\documentclass[prl,letterpaper,english,reprint,nofootinbib,aps,superscriptaddress,showpacs,showkeys]{revtex4-2}

\usepackage{babel,calc,amsmath,amsthm,amssymb,graphicx,subfigure,xcolor,comment}
\usepackage{mathdots}
\usepackage[T1]{fontenc}
\usepackage[unicode=true]{hyperref}
\usepackage{booktabs}
\usepackage{mathtools}
\usepackage{braket}
\usepackage{dsfont}
\usepackage{multirow}
\usepackage{multibib}

\definecolor{linkblue}{rgb}{0.1804, 0.1882, 0.5726}

\usepackage[normalem]{ulem}
\hypersetup{
	colorlinks=true,       		
	linkcolor=linkblue,          	
	citecolor=linkblue,             
	urlcolor=linkblue,          
}

\newtheorem*{theorem*}{Theorem}

\newtheorem*{corollary*}{Corollary}

\newtheorem*{lemma*}{Lemma}

\newtheorem*{proposition*}{Proposition}
\theoremstyle{definition}

\newtheorem*{definition*}{Definition}
\theoremstyle{remark}

\newtheorem*{remark*}{Remark}

\newif\ifdebug

\debugtrue

\ifdebug
\definecolor{zhliu}{rgb}{0.60, 0.12, 0.24}

\newcommand{\note}[1]{\textcolor{orange}{#1}}
\newcommand\delete{\bgroup\markoverwith{\textcolor{zhliu}{\rule[0.5ex]{2pt}{0.8pt}}}\ULon}

\else

\newcommand{\note}[1]{\ignorespaces}
\newcommand{\delete}[1]{\ignorespaces}

\fi

\begin{document}
	\renewcommand{\figurename}{Fig.}

	\title{Photonic Implementation of Quantum Information Masking}
	
	\author{Zheng-Hao~Liu}
	\affiliation{CAS Key Laboratory of Quantum Information, University of Science and Technology of China, Hefei 230026, People's Republic of China}
	\affiliation{CAS Centre For Excellence in Quantum Information and Quantum Physics, University of Science and Technology of China, Hefei 230026, People's Republic of China}
	
	\author{Xiao-Bin~Liang}
	\affiliation{School of Mathematics and Computer Science, Shangrao Normal University, Shangrao 334001, China}
	\affiliation{Quantum Information Research Center, Shangrao Normal University, Shangrao 334001, China}
	
	\author{Kai~Sun}
	\author{Qiang~Li}
	\author{Yu~Meng}
	\author{Mu~Yang}
	\affiliation{CAS Key Laboratory of Quantum Information, University of Science and Technology of China, Hefei 230026, People's Republic of China}
	\affiliation{CAS Centre For Excellence in Quantum Information and Quantum Physics, University of Science and Technology of China, Hefei 230026, People's Republic of China}

	\author{Bo~Li}
	\email{libobeijing2008@163.com}
	\affiliation{School of Mathematics and Computer Science, Shangrao Normal University, Shangrao 334001, China}
	\affiliation{Quantum Information Research Center, Shangrao Normal University, Shangrao 334001, China}
	
	\author{Jing-Ling~Chen}
	\email{chenjl@nankai.edu.cn}
	\affiliation{Theoretical Physics Division, Chern Institute of Mathematics, Nankai University, Tianjin 300071, People's Republic of China}
	
	\author{Jin-Shi~Xu}
	\email{jsxu@ustc.edu.cn}
	
	\author{Chuan-Feng~Li}
	\email{cfli@ustc.edu.cn}
	
	\author{Guang-Can~Guo}
	\affiliation{CAS Key Laboratory of Quantum Information, University of Science and Technology of China, Hefei 230026, People's Republic of China}
	\affiliation{CAS Centre For Excellence in Quantum Information and Quantum Physics, University of Science and Technology of China, Hefei 230026, People's Republic of China}

	\date{\today}
	
	\begin{abstract}
		Masking of quantum information spreads it over nonlocal correlations and hides it from the subsystems. It is known that no operation can simultaneously mask all pure states\,[Phys. Rev. Lett. \textbf{120}, 230501 (2018)], so in what sense is quantum information masking useful?
		Here, we extend the definition of quantum information masking to general mixed states, and show that the resource of maskable quantum states are far more abundant than the no-go theorem seemingly suggests. Geometrically, the simultaneously maskable states lays on hyperdisks in the state hypersphere, and strictly contains the broadcastable states. We devise a photonic quantum information masking machine using time-correlated photons to experimentally investigate the properties of qubit masking, and demonstrate the transfer of quantum information into bipartite correlations and its faithful retrieval. The versatile masking machine has decent extensibility, and may be applicable to quantum secret sharing and fault-tolerant quantum communication. Our results provide some insights on the comprehension and potential application of quantum information masking.
	\end{abstract}
	
	\maketitle
	
	\textit{Introduction.}---The distinctive nonclassicality of quantum mechanics establishes the pronounced discrepancy between quantum and classical information\,\cite{Bennett00}. 
	Especially, the celebrated quantum nonlocality\,\cite{EPR35} allude to the possibility of spreading information over the nonlocal correlation and hiding it from observers who only have access to some subsystems, namely, quantum information masking (QIM)\,\cite{Modi18}. 
	However, the postulates of unitarity and linearity\,\cite{Schrodinger26} of quantum theory impose severe limitations on QIM. In the seminal work\,\cite{Modi18}, unconditioned masking of all quantum states is deemed impossible. This assertion establishes a novel no-go theorem to paraphrase its conspicuous precedents like the interdicts against universal cloning\,\cite{Wootters82}, broadcasting\,\cite{Barnum96, Kalev08}, deleting\,\cite{Pati00, Samal11} and hiding\,\cite{Braunstein07} of an unknown state.
	
	Despite the handicap of universal implementation, QIM admits state-dependent\,\cite{xbliang19} and probabilistic realization\,\cite{bli19, msli19}, which is similar to other conditional quantum information tasks\,\cite{lmduan98, Bouwmeester02}.
	Extrapolations of masking into multipartite\,\cite{msli18}, multi-level\,\cite{fding20} scenarios and channel states\,\cite{Pereg20} have been proposed. QIM shows profound affiliation with quantum state discrimination\,\cite{gjtian15}, qubit commitment\,\cite{hklo97, Mayers97}, secret sharing\,\cite{Hillery99}, and fundamental principles like information conservation\,\cite{shlie20}. 
	Although significant progress has sprouted regarding QIM, its capability ultimately depends on the scale of maskable set, which is not yet determined.
	Moreover, given its intrinsically nonlocal feature, QIM on flying quanta has exceptional prospects of application in quantum communication. However, to our best knowledge, QIM has not been demonstrated in photonic experiments.
	
	The purpose of this Letter is twofold. We first provide a geometric description of QIM and prove the ``disk conjecture''\,\cite{Modi18}, that the maskable qudit states corresponding to any masker belong to some hyperdisks in the state hypersphere. This shows the copiousness of the maskable states, and allows us to identify the inclusion of the broadcastable states within it.
	Next, we devise a photonic masking machine capable of masking any disk in the qubit Bloch sphere, and give the recipe for qudit state masking. Assisted by the versatility of the machine, we experimentally confirm the geometric property of the maximally maskable set\,\cite{xbliang19, xbliang20}, and discuss the practical aspects of photonic QIM.
	Our results provide a systematic method for experimental studies of QIM, shed some lights on its applications as a novel quantum information processing protocol, and the connection between quantum information and nonlocality.

	\textit{Geometric characterization of QIM.}---We start from the formal definition of QIM for general quantum states.
	A qudit ``masker'' $\mathcal{U}$ is a linear isometry mapping a density matrix $\rho_s^A$ to a two-qudit state $\rho_s^{AB}$:
	\begin{align}
		\rho_s^{A} \rightarrow \rho_s^{AB}=\mathcal{U}\rho_s^A\otimes\ket{0}\bra{0}\mathcal{U}^\dagger, ~~s\in\{1, 2, \cdots\},
		\label{eq:isometry}
	\end{align}
	where $\ket{0}\bra{0}$ represents a blank state. We say that $\mathcal{U}$ ``masks'' the quantum information contained in a set $\Omega$ of density matrices $\left\{\rho_s^A\in\Omega\right\}$ if for all $s$, the marginal states of $\rho_s^{AB}$ for the two parties are respectively identical\,\cite{Modi18}. 
	
	To construct a geometric representation of a quantum state, we span a qudit state on the $SU(d)$ basis $\{\Lambda_i\}_{i=1}^{d^2-1}$. Explicitly, $\rho=I_d/d+\sum_{i=1}^{d^2-1}x_i\Lambda_i/2$ with $\sum_{i=1}^{d^2-1}x_i^2\leqslant r_d^2$, where $x_i$ represents the coefficients, and $r_d$ is determined by the dimension $d$\,\cite{Mahler95, Kimura03}. Consequently, every qudit state corresponds to a unique point in a hypersphere, which is analogous to the Bloch sphere for qubits.
	Based on this representation, we directly compare the coefficients of the local states after masking to bound the maskable sets, and use the impossibility of universal masking\,\cite{Modi18} to prove the following result:
	\vspace{3pt}
	
	\textit{Theorem 1.}---The maskable set corresponds to any linear qudit isometry lays on some hyperdisk $\mathcal{D}$.
	\vspace{3pt}
	
	The above Theorem first appears in~\cite{Modi18} as a conjecture, and its proof goes to Section IB of the Supplementary Material\,\cite{SM} (SM).
	Because the Euclidean dimension of a hyperdisk is smaller than a hypersphere by only 1, Theorem 1 shows the possibility of masking a large class of quantum states and the potential of QIM in quantum information tasks.
	It also inspires us to identify the relationship between the sets of broadcastable and maskable quantum states. Specifically, because noncommuting mixed states cannot be broadcast\,\cite{Barnum96}, and commuting mixed states are simultaneously maskable (cf. SM\,\cite{SM}, Section IC), we have:
	\vspace{3pt}
	
	\textit{Theorem 2.}---Any qudit broadcastable set is a proper subset of some qudit maskable sets. Moreover, The qudit maskable set can have nonzero measure in $d$-dimensional Euclidean space.
	\vspace{3pt}
	
	The implication of our results can be clearly visualized in the qubit case using the Bloch sphere representation. For simplicity, we interchangeably denote a state $\rho=\left(I_2+x\sigma_x+y\sigma_y+z\sigma_z\right)/2$ using its $SU(2)$ expansion $\rho:=(x,y,z)$. 
	A qubit disk containing the reference state $\rho_0=(x_0,y_0,z_0)$ can be expressed in a parametric form: 
	\begin{align}
		\mathcal{D}_\alpha^\theta (\rho_0) &= \{ \rho: x\sin \alpha \cos \theta + y\sin \alpha \sin \theta + z\cos \alpha = c \} ,
	\end{align}
	with $c=x_0\sin\alpha\cos\theta+y_0\sin\alpha\sin\theta+z_0\cos\alpha,\, \alpha\in[0,\pi]$ and $\theta\in[0,2\pi]$. 
	In comparison, the geometric form of the broadcastable set is a line segment through the center of the Bloch sphere, so the dimensions of the disk and line segment conforms Theorem 2, and the resource in the maskable set is far more abundant than the broadcastable set. Notably, the qubit isometry $\mathcal{U}_\alpha^\theta$ capable of masking the disk $\mathcal{D}_\alpha^\theta (\rho_0)$ always exists (cf. SM\,\cite{SM}, Section IB), and can be devised and reliably implemented on the photonic architecture, as will be elucidated in the following section.
	
	\begin{figure}[htbp]
		\centering
		\includegraphics[width = .48 \textwidth]{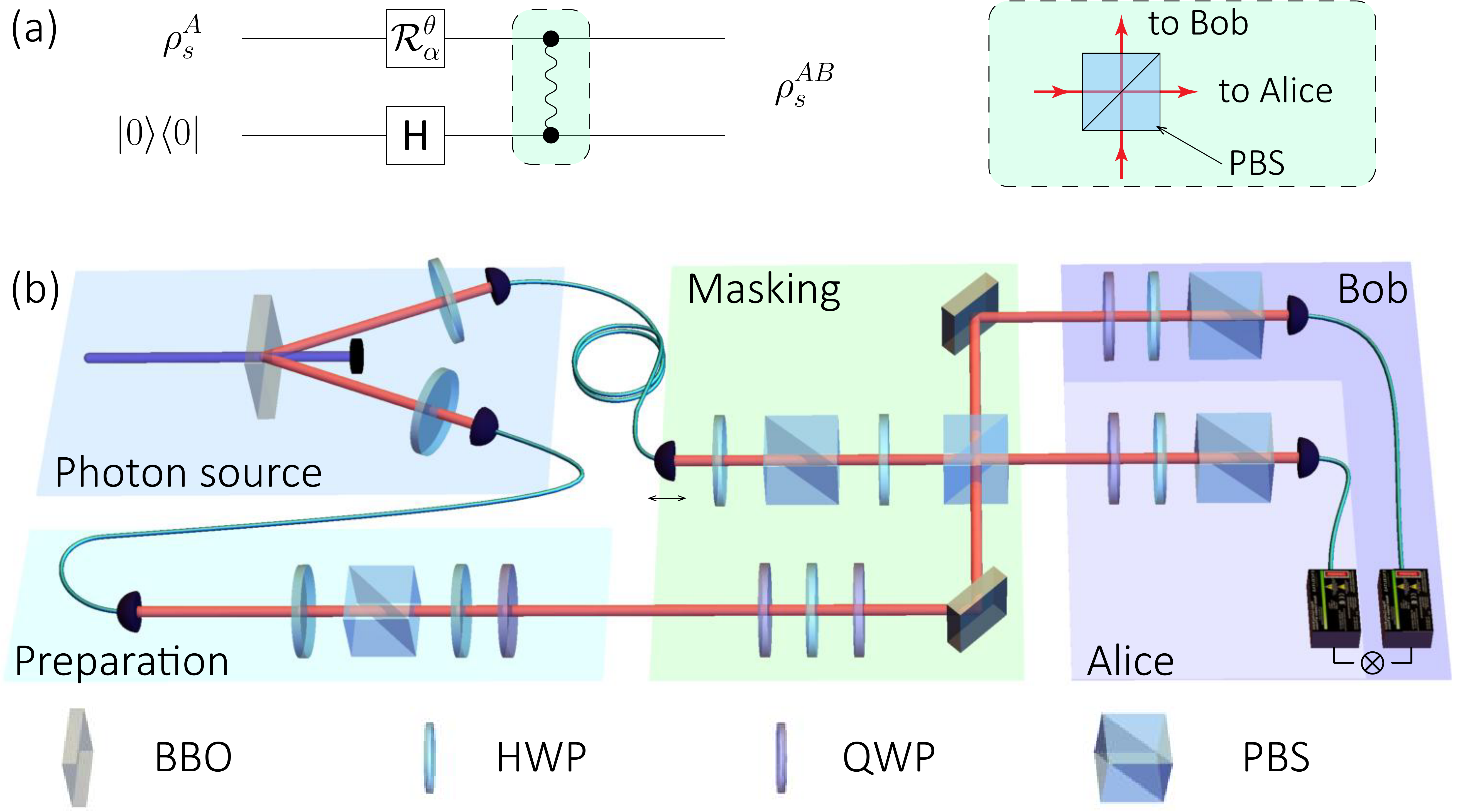}
		\caption{Photonic qubit masking machine. (a) The quantum circuit of the masking machine. $\mathcal{R}_\alpha^\theta$ and $\mathrm{H}$ represents an arbitrary $SU(2)$ rotation and the Hadamard gate, respectively. The tilde line in the rounded rectangle denotes the photon fusion gate, with the detailed implementation using the polarizing beam splitter (PBS) illustrated in the right subpanel. (b) Experimental configuration. See the main text for details.
		}
		\label{fig:setup}
	\end{figure}

	\textit{A photonic masking machine.}---The polarization degree of freedom of the photons is a natural courier of qubit information. Specifically, the correspondences $\ket{H}\leftrightarrow\ket{0}, \ket{V}\leftrightarrow\ket{1}$ link the isomorphic Hilbert spaces of a photon's polarization state and a qubit state, with $\ket{H}$ and $\ket{V}$ denoting the horizontal and vertical polarization of the photon, respectively. 
	We exploit the photon fusion gate\,\cite{Browne05} to construct a class of maskers $\mathcal{U}_0^0$, which is further promoted to arbitrary parameters $\mathcal{U}_\alpha^\theta$ using additional wave plates. 
	
	The quantum circuit of the masking machine is illustrated in Fig.\,\ref{fig:setup}a. The photon fusion gate performs a two-photon interference on a polarizing beam splitter (PBS), 
	and is conditioned on two-photon coincidence detection at two different output ports. This effectively casts an entangling projector $\ket{HH}\bra{HH}-\ket{VV}\bra{VV}$ onto the input photons\,\cite{lmduan06}. Given an auxiliary photon initialized in the $\ket{D}=(\ket{H}+\ket{V})/\sqrt{2}$, the behavior of the fusion gate on the target photon is equivalent to the masking isometry $\mathcal{U}_0^0$ up to a re-normalization (cf. SM\,\cite{SM}, Section IIA). More explicitly, applying the fusion gate on a qubit state $\ket{\psi}=\cos\delta\ket{H}+\sin \delta e^{i\phi}\ket{V}$ yields $\ket{\psi}\otimes\ket{D}\to \cos\delta\ket{HH}-\sin\delta e^{i\phi}\ket{VV}$. Because the phase factor $\phi$ does not appear in either of the marginal states, all the states with the same $\delta$ can be masked, and they falls on the disk $\mathcal{D}_0^0=\{\rho:z=\cos\delta\}$.
	
	Using this photonic masking machine, an agent (Alice) can conceal some quantum information into the bipartite correlation between her photons and the ones held by another agent (Bob). The experimental setup is illustrated in Fig.\,\ref{fig:setup}b. An ultraviolet laser with a central wavelength of 400nm is used to pump a type-II phase-matched $\beta$-barium borate (BBO) crystal to generate photon pairs in product polarization states via the spontaneous parametric down-conversion process. 
	The photon pairs are collected by two single-mode fibers (SMFs), with one of the photons sent to Alice for initial state preparation using a half-wave plate (HWP) and a quarter-wave plate (QWP), and the other photon directly fed into the masking machine, where its polarization is rotated to $\ket{D}$.
	
	Subsequent to initial state preparation, Alice inputs her photon to the masking machine, which first undergoes a polarization rotation induced by a HWP sandwiched by two QWPs. This rotation maps the arbitrary disk to be masked to a latitudinal plane in the Bloch sphere\,\cite{Englert01}.
	The two photons are then interfered on a PBS, with their trajectory and arrival time carefully aligned to ensure the proper overlapping of the spatial and temporal wavefunctions.
	The output photon from one port is kept by Alice, and the other is sent to Bob. The polarization states of the photons are later analyzed by a PBS preceded by a QWP and a HWP. Finally, the photons are again collected by two SMFs and sent to single-photon avalanche detectors for coincidence counting.
	
	\begin{figure}[t]
		\centering
		\includegraphics[width = .96 \columnwidth]{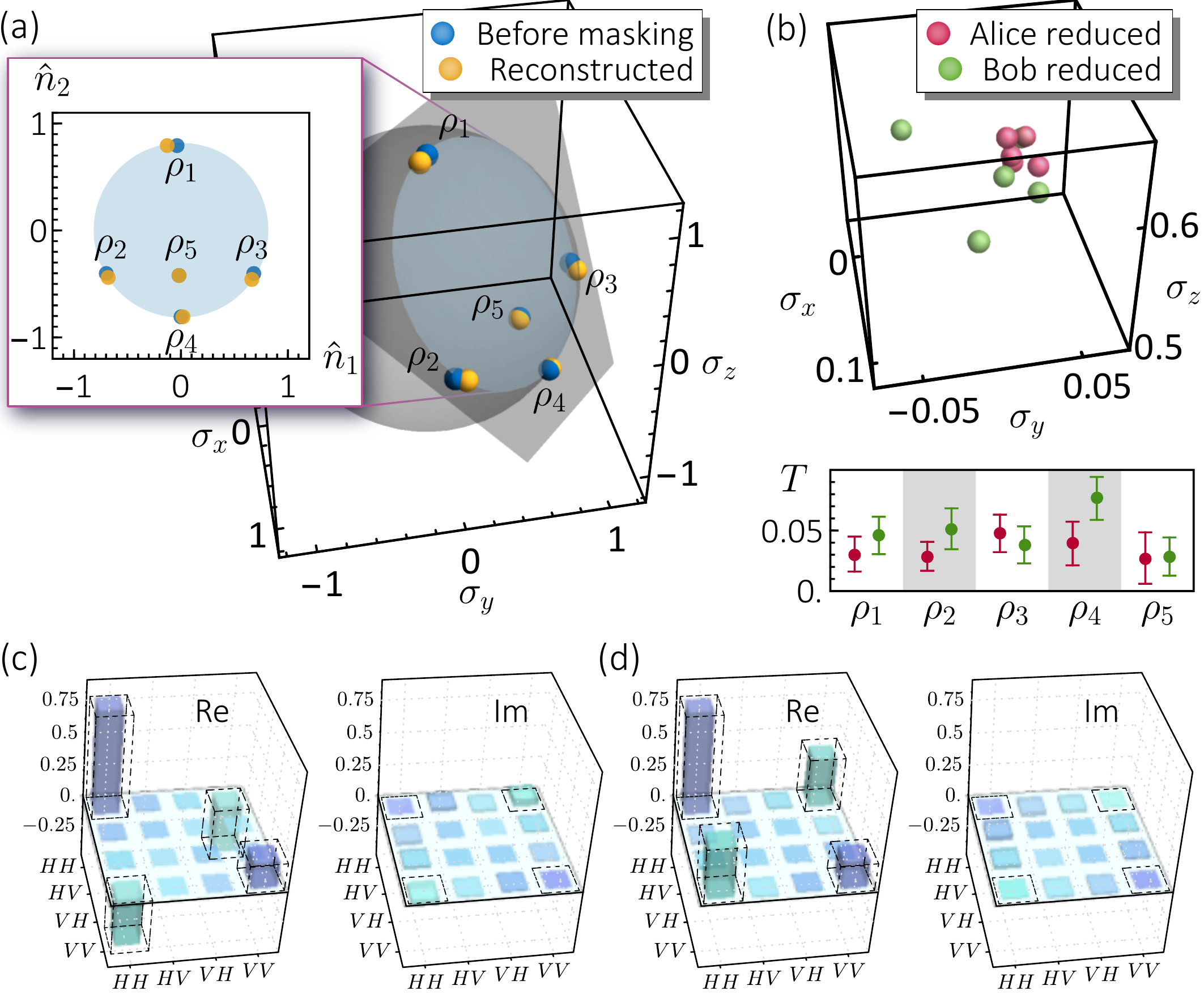}
		\caption{Masking of a qubit disk. 
			(a) The maskable disk is the intersection of the plane $x+y+z=1$ and the unit ball. Pure state $\rho_1\sim\rho_4$ and mixed state $\rho_5$ fall on the disk. The blue and orange points denote the initial states deduced from quantum state tomography and the reconstructed final states, respectively. The inset shows a projection of these states on the maskable disk, and the axes are defined as $\hat{n}_1=(-\hat{x}+\hat{y})/\sqrt{2}$ and $\hat{n}_2=(-\hat{x}-\hat{y}+2\hat{z})/\sqrt{6}$.
			(b) Experimentally determined marginal states after masking, and the trace distance from their theoretical values. 
			(c) and (d) The density matrix of the bipartite state resulted from masking $\rho_1$ and $\rho_2$. Solid and dashed bars denote the experimental values and theoretical predictions, respectively.
		}
		\label{fig:demo}
	\end{figure}

	To exemplify the aptitude of the masking machine, we experimentally realize $\mathcal{U}_\vartheta^{\pi/4}$ and mask the disk $\mathcal{D}_\vartheta^{\pi/4}(\rho_1)$, with $\vartheta=\arctan\sqrt{2}$. The disk passes through $\rho_1=\ket{H}\bra{H}=(0,0,1),\rho_2=\ket{D}\bra{D}=(1,0,0)$, and $\rho_3=\ket{L}\bra{L}=(0,1,0)$ with $\ket{L}=(\ket{H}+i\ket{V})/\sqrt{2}$. The form of $\mathcal{U}_\vartheta^{\pi/4}$ requires the orientation of the three cascaded wave plates in the masking machine set to $58.18^\circ, 0^\circ$ and $64.66^\circ$. 
	The masker is also applicable for the other states on the disk, and here we test the cases of the pure state $\rho_4=\left(2/3,2/3,-1/3\right)$ and the mixed state $\rho_5=\left(1/2,1/2,0\right)$, with the latter prepared using the temporal-mixing technique\,\cite{Dakic12}. 
	
	The experimental initial states in a Bloch sphere (blue dots) are shown in Fig.\,\ref{fig:demo}a. To retrieve the masked information, we numerically apply the inverse isometry $\mathcal{U}^{-1}$ on the reconstructed bipartite density matrix. The final state is also shown in Fig.\,\ref{fig:demo}a (orange dots) for comparison. The reconstruction achieved a mean fidelity of 99.87\%, and the average total absolute spectra error is $3.72\times10^{-2}$. 
	The effect of masking can be further reflected by the marginal states of Alice and Bob. See Fig.\,\ref{fig:demo}b, they almost completely overlap, with the average trace distance $T\left(\rho_i,\rho_j\right)=\frac{1}{2}\|\rho_i-\rho_j\|_1$\,\cite{QCQI} of Alice's and Bob's marginal states being $1.55\times10^{-2}$ and $4.06\times10^{-2}$, respectively. 
	On the other hand, joint measurements on two photons show that the average fidelity of the masking-resulted states with respect to the theoretical predictions is $97.70\%$. Two instances of reconstructed bipartite density matrices are shown in Fig.\,\ref{fig:demo}c and \ref{fig:demo}d, in accord with their theoretical values. Overall, the quantum information has almost completely retreated from the local marginal states, and is faithfully kept in the bipartite correlation.
	
	\begin{figure}[b!]
		\centering
		\includegraphics[width = .48 \textwidth]{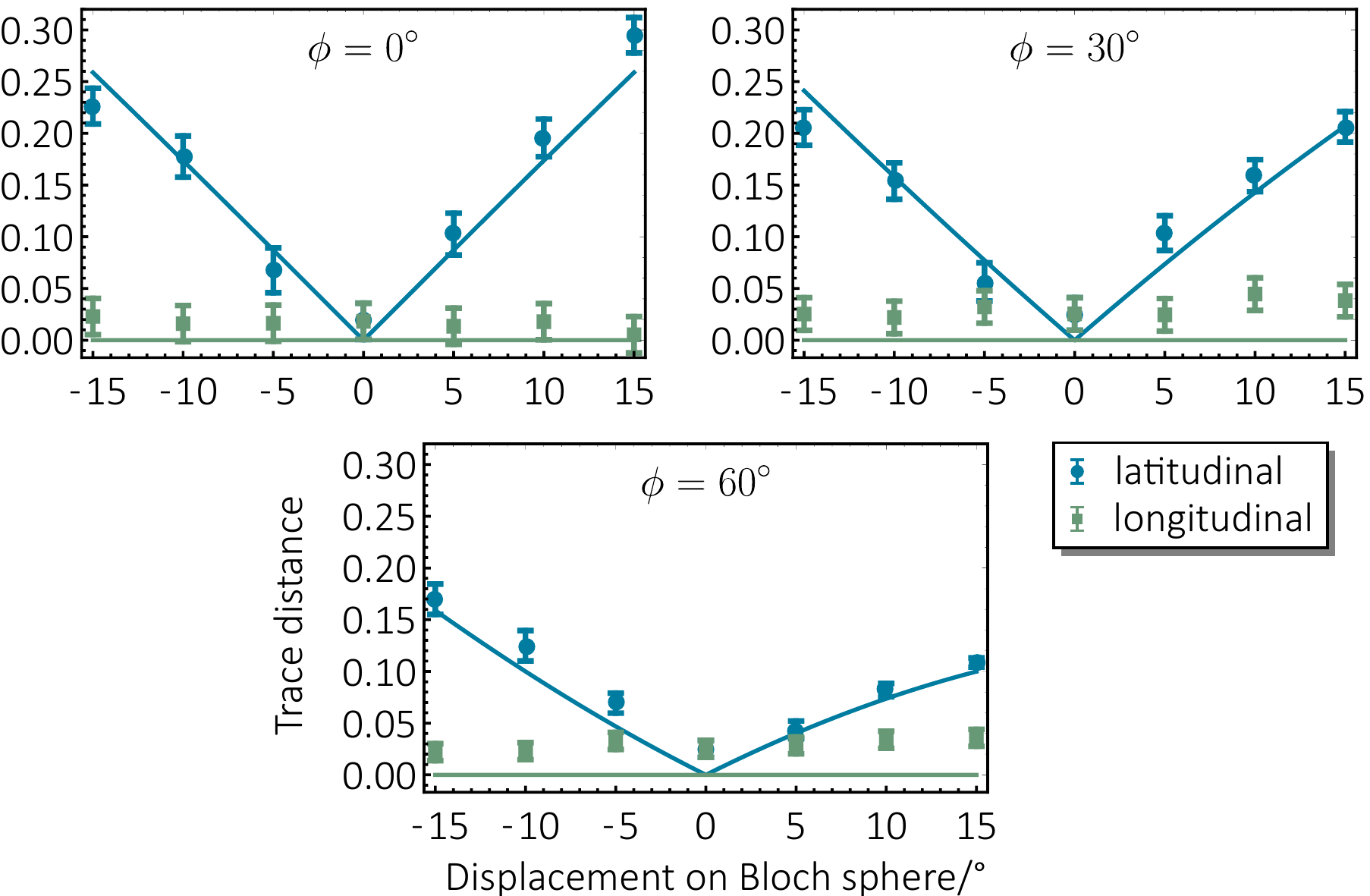}
		\caption{Zero measure of maskable sets. The maskable disks are encircled by the line of constant latitudes $\phi$ on the Bloch sphere. 
			Each plot shows the trace distance of Bob's marginal density matrices from the reference state, when Alice slightly shifts the initial state away from a reference point on different latitudes. The cases of displacement along a parallel or a meridian on the Bloch sphere are plotted respectively in green and cyan points (experimental data) and curves (theoretical values). The error bars correspond to the $1\sigma$ standard deviation, deduced from a Poissonian counting statistics.
		}
		\label{fig:measure}
	\end{figure}
	
	Furthermore, we observe the zero Haar measure of the maskable set, that is, the thickness of the maskable disk is infinitesimal. We verify this property on the masker $\mathcal{U}_0^0$, realized by setting the orientations of all the three wave plates in the masking machine to $0^\circ$. The corresponding maskable disks are every latitudinal plane on the Bloch sphere. 
	Experimentally, the reference states $\psi_0=\sin(\phi/2)\ket{H}+\cos(\phi/2)\ket{V}$ on latitude $\phi$, with $\phi$ setting to $0^\circ,30^\circ$ and $60^\circ$, are selected. We then prepare some states shifted from $\psi_0$ along a parallel or a meridian, cast $\mathcal{U}_0^0$, and take tomography on these states. The marginal states of Bob, $\rho^B$, are obtained from the reconstructed bipartite density matrices, and compared with the theoretical value, $\rho_0^B=\mathrm{Tr}_A\left(\mathcal{U}_0^0\,\ket{\psi_0}\bra{\psi_0}\otimes\ket{0}\bra{0}\mathcal{U}_0^{0\dagger}\right)$ to determine the result of masking.
	Our results in Fig.\,\ref{fig:measure} shows that when the shift is along a parallel, $\rho^B$ remains invariant, which can be revealed by the vanishing $T(\rho^B,\rho_0^B)$. Consequently, the state still belongs to the maskable set. However, a shift along a meridian always induces nonzero $T(\rho^B,\rho_0^B)$ regardless of $\phi$, indicating failure of qubit masking because extra information is transmitted to the marginal state of Bob. 
	We note that the residual error in the longitudinal data is mainly due to the imperfect overlapping of the two-photon wavefunction on the PBS, which is caused by the spectral difference between the two type-II parametric photons\,\cite{Grice97}.
	
	\textit{Discussion.}---Our photonic masking machine is based on the qubit photon fusion gate, but the adopted method here is also capable of masking qudit states. To this end, we can encode each digit of the qudit on a qubit, akin to the practice in the quantum factoring algorithm\,\cite{Shor94, Obrien11}, and mask every qubit independently. A rigorous account for the qudit state masking is deferred to Section IIB of SM\,\cite{SM}. Moreover, because the fusion gate only requires time-correlated photons, the masking machine will work for not only the parametric photon pairs, but also for paused weak coherent light as one input mode, providing the other input mode is genuine single-photon with the same wavelength and repetition rate (cf. SM\,\cite{SM}, Section IIC). These features suggest that QIM may be useful in bright single-photon source-based quantum information processing tasks.
	
\begin{figure}[t]
		\centering
		\includegraphics[width = .96 \columnwidth]{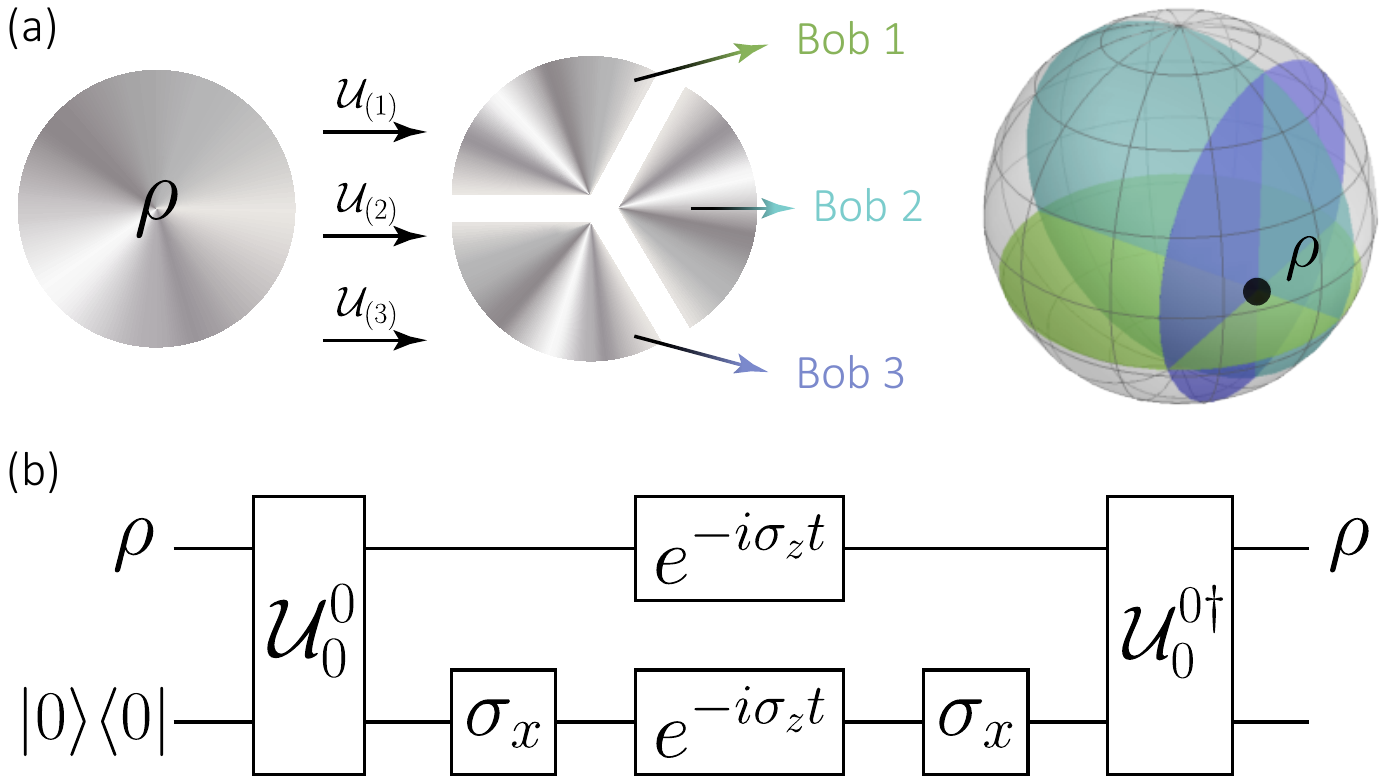}
		\caption{Application of QIM. 
			(a) Sharing of a secret state among multiple recipients. Alice masks a secret quantum state $\rho$ using three maskers $\mathcal{U}_i$, and send one resulting marginal state to $\text{Bob}_i$, respectively. From the information of the marginal states and the masker informed by Alice, every Bob independently interpret the possible disk that contains the masked states. The original state is pinned at the intersection of the three disks. 
			(b) Protecting quantum information in a noisy channel. Given only access to the quantum channels with phase errors $\exp \left(-i\sigma_zt\right)$, a qubit can still be correctly transmitted using QIM based on the masker $\mathcal{U}_0^0$.
		}
		\label{fig:discuss}
	\end{figure}

	QIM has some practical merits in addition to the theoretical significance. As a proof-of-concept application, we utilize the infinitely many $d$-dimensional maskable set to experimentally demonstrate quantum secret sharing\,\cite{Hillery99, Cleve99, Tittel01, Lance04, Williams19}. As is shown in Fig.\,\ref{fig:discuss}a, Alice masks a quantum state using the masking machine tuned to three different maskers $\mathcal{U}_{(i)}$, and sends each resulting qubit to a recipient, $\text{Bob}~i$, respectively. The Bobs can only use their marginal state to restrict the masked state onto a disk, and have to cooperate together and comparing their results to reveals the concealed information: upon comparison, the three disks will intersect at a single point in the Bloch sphere. We explain the experimental details, the security of the secret sharing, and its application in image reconstruction in section IID of SM\,\cite{SM}. Owing to the geometric representation of the maskable set, our scheme does not rely on entanglement between the recipients and suffer from decoherence. It also works smoothly for mixed state. Moreover, it requires no Pauli-type correction at the receiver's side, thus is applicable even when the sender has no access to classical communication after masking.
	
	The quantum information processing protocols based on QIM also have intimate connection with fault-tolerance. Because the quantum information is transferred to the nonlocal correlation after masking, it acquires additional resilience to some common-mode noise. An example is shown in Fig.\,\ref{fig:discuss}b. Suppose we are given access to two identical, noisy quantum channels, which will apply a random phase error $e^{-i\sigma_zt}$ on the input state. To correctly transmit a qubit state, we can mask it with $\mathcal{U}_0^0$, and apply $\sigma_x$ on the auxiliary qubit before sending the two qubits through the channel. After the transmission, the original quantum state can be recovered by applying $\sigma_x$ again on the auxiliary qubit and using the inverse isometry $\mathcal{U}_0^{0\dagger}$. Since the protocol resembles the celebrated spin echo effect (with the noise applied simultaneously on two quanta), it is plausible to expect that it will find further applications in the near future.
	
	Going beyond the no-go theorem, we have provided both the geometric properties of QIM, and the recipe for implementing QIM on photonic architectures. We have demonstrated QIM on our photonic qubit masking machine to meticulously test the theoretical predictions, and show that quantum information can be concealed in and restored from nonlocal correlations. The masking machine is extensible, versatile, and applicable in tasks like quantum secret sharing and quantum communication. This work deepens our comprehension of the tie between QIM and the basic axioms of the quantum nature, and sheds lights on its future applications in quantum information processing.

	\let\oldaddcontentsline\addcontentsline
	\renewcommand{\addcontentsline}[3]{}
	\textit{Acknowledgments.}---
	%
	This work was supported by
	National Key Research and Development Program of China (Grants Nos.\,2016YFA0302700, 2017YFA0304100),
	the National Natural Science Foundation of China (Grants
	Nos.\,61725504, U19A2075,
	12065021,
	61805227, 61975195,
	11774335, and 11821404),
	Key Research Program of Frontier Sciences, CAS (Grant No.\,QYZDY-SSW-SLH003),
	Science Foundation of the CAS (Grant No.\,ZDRW-XH-2019-1),
	the Fundamental Research Funds for the Central Universities (Grant No.\,WK2470000026, No.\,WK2030380017),  
	Anhui Initiative in Quantum Information Technologies (Grants No.\,AHY020100, and No.\,AHY060300).
	J.L.C. was supported by the National Natural Science Foundations of China (Grant No.\,11875167).
	X.B.L. was supported by the Natural Science Foundation of Jiangxi Province (Grant No.\,20202BAB201010).
	
	Z.H.L., X.B.L and K.S. contributed equally to this work.

	\onecolumngrid\phantom{123}
	\let\addcontentsline\oldaddcontentsline 
	

	\clearpage
	\newpage
	\setcounter{page}{1}
	\appendix
	\onecolumngrid
	
	\setcounter{equation}{0}
	\setcounter{figure}{0}
	\renewcommand{\theequation}{S\arabic{equation}}
	\renewcommand{\thefigure}{S\arabic{figure}}
	\renewcommand{\appendixname}{Section}

	\begin{center}
		\textbf{\large Supplementary Material for \\ ``Photonic Implementation of Quantum Information Masking''}
	\end{center}
	
	\setcounter{equation}{0}
	\setcounter{figure}{0}
	\renewcommand{\theequation}{S\arabic{equation}}
	\renewcommand{\thefigure}{S\arabic{figure}}

	In this Supplementary Material, we give the proofs for the propositions in the main text. We first introduce the geometric representation of a qudit state. Exploiting this representation, we provide a geometric feature of the maximal maskable set in an arbitrary $n$-dimensional qudit Hilbert space, and subsequently derive the form for qubit masking, which correspond to $n=2$. Finally, we prove that any commuting, mixed states set can be simultaneously masked, and the set of quantum states that is broadcastable is a proper subset of one that is maskable.
	
	\tableofcontents
	
	\section{Theoretical considerations}
	
	In the theoretical part of the Supplementary Material, we give the proofs for the propositions in the main text. We first introduce the geometric representation of a qudit state. Exploiting this representation, we prove the disk conjecture (Theorem 1) regarding the geometric form of the maximal maskable set for $d$-dimensional qudit states, and subsequently prove the inverse proposition in qubit case, that a two-dimensional quantum disk must also be maskable. We also explicitly derive the form of isometry for qubit masking. Finally, we prove (Theorem 2) that any qudit broadcastable set is a proper subset of some qudit maskable sets. Moreover, the qudit maskable set has nonzero measure in $d$-dimensional Euclidean space.
	
	\subsection{Geometric representation of qudit}
	
	An arbitrary qudit state $\rho$ can be expressed as follows using the $SU(d)$ basis $\{\Lambda_i\}^{d^2-1}_{i=1}$, where $\Lambda_i$ are orthogonal eigenstates satisfying $\Lambda_i^+=\Lambda_i$, $\mathrm{Tr}(\Lambda_i)=0$, and $\mathrm{Tr}(\Lambda_i\Lambda_j)=2\delta_{ij}$. Namely\,\cite{Mahler95, Kimura03},
	
	\begin{eqnarray}
		\rho=\frac{1}{d}I_d+\frac{1}{2}\sum^{d^2-1}_{i=1}x_i \Lambda_i,~~~ \sum^{d^2-1}_{i=1}x^2_i\leq r_d^2,~~~ x_i\in \mathbb{R},
	\end{eqnarray}
	where $r_d$ satisfies $${\frac{2}{d(d-1)}}\leq r_d^2 \leq {\frac{2(d-1)}{d}}.$$
	
	The definition of the hyperdisk is crucial for the comprehension of quantum information masking. A hyperdisk $\mathcal{D}$ is defined as the intersection of a hypersphere and a hyperplane in the Euclidean space. Namely,
	
	\begin{align}
		\mathcal{D}=\{(x_1,x_2,...,x_d):\sum^d_{i=1}x^2_i\leq r^2_d,\sum^d_{i=1}a_ix_i+b=0,a_i \mathrm{~not~ all~ be~ zero}\}.
	\end{align}
	
	\subsection{General form of maskable set in arbitrary dimensions}
	
	\textit{Theorem 1.}---The maximal maskable set for an arbitrary $d$-dimensional qudit state lays on a hyperdisk in $d^2-1$ dimensional Euclidean space.
	
	\begin{proof}---
		Let $\rho$ be a density matrix on the Hilbert space $\mathcal{H}_A$, which is the mixed states to be masked, and $\mathcal{U}$ is the masker, a unitary operator acting on $\mathcal{H}_A\otimes\mathcal{H}_B$ as: $\mathcal{U}\rho\otimes|0\rangle\langle0|\mathcal{U}^\dag=\rho^{AB}.$
		Further, let the marginal state $\rho_A=\mathrm{Tr}_B(\rho^{AB})$, then
		
		\begin{align}
			\rho_A(k,l)=\sum^{d^2-1}_{i=1}a^{(kl)}_ix_i+b^{(kl)}, k,l=1,2,...,d.
		\end{align}
		
		where $\rho_A(k,l)$ is the $k$-th row and $l$-th column element of matrix $\rho_A$, $a^{(kl)}_i,b^{(kl)}\in \mathbb{C}$. 
		A given subset of quantum states in $\mathcal{H}_A$ is maskable iff. $\rho_A(k,l)$ for $\rho_A$, and similarly $\rho_B(k,l)$ for $\rho_B$, are constant, namely, $\exists~ h^{(kl)}_X\in \mathbb{C},~ \rho_X(k,l)\equiv h^{(kl)}_X$, $X=A~ \mathrm{or}~ B$. 
		Obviously, for a masker $\mathcal{U}$, if all $\rho_X(k,l)$ are constant for all $\rho$, then masking of all quantum qudit pure states is possible, which contradicts the known results, so $a^{(kl)}_i,i=1,2,...,d$ can't all be zero. In other words, there exists nontrivial hyperplane $\mathrm{Re}\,\rho_X(k,l)=\mathrm{Re}\,h^{(kl)}_X$ or $\mathrm{Im}\,\rho_X(k,l)=\mathrm{Im}\,h^{(kl)}_X$, which contains all of the maskable states corresponding to the masker $\mathcal{U}$. So the maximal maskable set of states with respect to $\mathcal{U}$ always lays on a hyperdisk in $d^2-1$ dimensional space.
		
	\end{proof}
	
	Note that for the qubits, we have $d = 2, r_d \equiv1$. In this case, the map between the set of points in the Bloch sphere and the set of qubit states is a bijection, so the maximal maskable set of qubit states with respect to $\mathcal{U}$ is on a disk in the Bloch sphere. Moreover, an arbitrary disk passing through the point $\rho_0=(x_0,y_0,z_0)$ in the Bloch sphere can always be represented as $\mathcal{D}^\theta_\alpha(\rho_0)$.
	We now prove that any qubit disk $\mathcal{D}^\theta_\alpha(\rho_0)$ can be masked. Let's define the qubit masker $\mathcal{U}^\theta_\alpha$, with
	\begin{align}
		\label{eqs:uat}
		\mathcal{U}_\alpha^\theta &= \begin{pmatrix}\cos(\alpha/2) & 0 & e^{-i\theta}\sin(\alpha/2) & 0 \\ 0 & \cos(\alpha/2) & 0 & e^{-i\theta}\sin(\alpha/2) \\ 0 & \sin(\alpha/2) & 0 & -e^{-i\theta}\cos(\alpha/2) \\ \sin(\alpha/2) & 0 & -e^{-i\theta}\cos(\alpha/2) & 0\end{pmatrix}.
	\end{align}
	We apply the masker on the initial state to find the final state resulted by masking, which reads:
	\begin{align}
		\rho^{AB} = \mathcal{U}^\theta_\alpha\rho\otimes|0\rangle\langle0|{\mathcal{U}^\theta_\alpha}^\dag.
	\end{align}
	Then, direct calculation shows that the marginal density matrices for any $\rho\in\mathcal{D}_\alpha^\theta(\rho_0)$ are identical:
	\begin{subequations}
		\begin{align}
			\rho_A\equiv \mathrm{Tr}^B[\rho_{AB}] &= \frac{1}{2}(1+c)|0\rangle\langle0|+\frac{1}{2}(1-c)|1\rangle\langle 1|, \\
			\rho_B\equiv \mathrm{Tr}^A[\rho_{AB}] &=  \frac{1}{2}(1+c)|0\rangle\langle0|+\frac{1}{2}(1-c)|1\rangle\langle 1|, \\
			\text{with}~~~ c &= x_0\sin\alpha\cos\theta+y_0\sin\alpha\sin\theta+z_0\cos\alpha. \nonumber
		\end{align}
	\end{subequations}
	
	Furthermore, we comment on the measure of the maskable set. Because the volume of the Bloch sphere is $\frac{4\pi}{3}$ and the maximal maskable set is a disk with a volume of zero, the measure of the maskable set corresponding to all qubit states is zero. This conclusion also holds for qudit states, because the volume of all qudit states is greater than the volume of the ball of with $r = \sqrt{\frac{2}{d(d-1)}}$, and the volume of hyperdisk is zero because it is restricted in an $d^2-1$ dimensional space. This lead to the conclusion that a maskable set corresponding to any isometry has zero measure.

	\subsection{Strict inclusion of broadcastable set in maskable set}
	
	In this section, we consider the relationship between the set of the maskable and broadcastable qudit states. Here, we show that the latter set is a proper subset of the former one. Our start point is that the density matrices of the broadcastable states commute\,\cite{Barnum96}, and the proof is based on the following observation.
	
	\textit{Lemma 1.}---Commuting mixed states set can be simultaneously masked.
	
	\begin{proof}   
		The density operators of mixed states are represented by Hermitian matrices. As such, the density operators of commuting mixed states can be simultaneously diagonalized. Suppose the spectrum decomposition of $\rho^A_s$ is:
		
		\begin{align}
			\label{eq:bcable}
			\rho^A_s=\sum^{d}_{i=1}p_i|i\rangle\langle i|, ~\mathrm{for~ all}~ \sum^{d}_{i=1}p_i=1.
		\end{align}
		
		The equation (\ref{eq:bcable}) represents a hyperplane in $d$-dimensional Euclidean space, and have zero volume measure. Let's construct a Vandermonde matrix $(a_{kl})_{d\times d}$ of unit roots, namely
		
		\begin{eqnarray}
			(a_{kl})_{d\times d}=\left(\begin{array}{ccccc}
				1 & 1 & 1 & \cdots& 1\\
				x_1 & x_2 & x_3 & \cdots& x_d\\
				x^2_1 & x^2_2 & x^2_3 & \cdots& x^2_d\\
				\vdots & \vdots & \vdots & \ddots& \vdots\\
				x^{d-1}_1 & x^{d-1}_2 & x^{d-1}_3 & \cdots& x^{d-1}_d\\ \end{array} \right
			),
		\end{eqnarray}
		
		where $x_i$ are the roots of $x^d = 1$. One may verify that the inner product of the two different row vectors of this matrix $(a_{kl})_{d\times d}$ is zero. Let we denote $|\Psi_k\rangle=\frac{1}{\sqrt{d}}\sum^{d}_{l=1} a_{kl}|l\rangle|l\rangle$, then ${\{|\Psi_k\rangle\}^d_{k=1}}$ is a set of orthonormal bases.
		Also, we choose the isometric linear operator $\mathcal{M}$ acting on the base $\{|k\rangle\}^d_{k=1}$ in the following way,
		\begin{eqnarray}
			\label{eqs:vdm}
			& |k\rangle \rightarrow |k\rangle|1\rangle\rightarrow \Psi_k,~ k=1,\ldots,d.\nonumber
		\end{eqnarray}
		
		From the definition of the isometry, $\mathcal{M}$ can always be dilated to an unitary operator $\mathcal{U}$ on the space $\mathcal{H}_A\otimes \mathcal{H}_B $, that is, there exist a unitary operator $\mathcal{U}$,  $\mathcal{U}|k\rangle|0\rangle=\mathcal{M}|k\rangle=|\Psi_k\rangle$.  Note that $\mathcal{U}\rho^A_s\otimes|0\rangle\langle0|\mathcal{U}^\dag=\rho^{AB}_s$. Direct calculation of $\rho_A$ and $\rho_A$ yields the marginal density matrices 
		\begin{eqnarray}\rho_A=\rho_B=\frac{1}{d} \sum^{d}_{k=1}\sum^{d}_{i=1}p_i \parallel a_{ik}\parallel|k\rangle\langle k|\equiv \frac{1}{d}\sum^{d}_{k=1}|k\rangle\langle k|.\nonumber\end{eqnarray}
		Where $\rho_A=\rho_B$ is guaranteed by the symmetric form of the isometry. We see that the eigenvalues $p_i$ cease to appear in the final reduced density matrices, so the masking succeed.
	\end{proof}
	
	We now proceed to prove the Theorem 2 in the main text.
	
	\textit{Theorem 2.}---Any qudit broadcastable set is a proper subset of some qudit maskable sets. Moreover, a qudit maskable set has nonzero measure in $d$-dimensional Euclidean space.
	
	\begin{proof}
		Let 
		\begin{align}
			\ket{\psi_0} = \left(\frac{d-1}{d}+\frac{1}{d}i\right)\ket{1} + \sum_{k=2}^d\left(-\frac{1}{d}+\frac{1}{d}i\right)\ket{k},
		\end{align}
		where $i$ is the unit imaginary number. Applying the isometry $\mathcal{M}$ defined in (\ref{eqs:vdm}) on $\ket{\psi_0}$, we find $\mathrm{Tr}_A(\mathcal{M}\ket{\psi_0}\bra{\psi_0}\mathcal{M}^\dagger) = \mathrm{Tr}_B(\mathcal{M}\ket{\psi_0}\bra{\psi_0}\mathcal{M}^\dagger) = \frac{1}{d}\sum^{d}_{k=1}|k\rangle\langle k|$, and thus $\ket{\psi_0}\bra{\psi_0}$ and $\rho_s^A$ is simultaneously maskable by $\mathcal{M}$. 
		By the linearity of the trace operation, the set of quantum states living in $d+1$-dimensional real space $\mathbf{P}=\{\rho\,\vert\,\lambda_0 \ket{\psi_0}\bra{\psi_0} + \sum_{i=1}^d \lambda_k \ket{k}\bra{k}, ~\sum_{k=0}^d \lambda_k=1\}$ is maskable by $\mathcal{M}$. It is straightforward to check that for $\prod_{i=0}^d\lambda_i\neq0$ the states $\rho\in\mathbf{P}$ do not mutually commute, so they are predominantly not broadcastable. 
		Moreover, as the qudit maskable set lives on the $d+1$-dimensional hyperplane which has nonzero volume measure on the $d$-dimensional Euclidean space, so the dimension of the set of the maskable quantum states is at least 1 plus which of the broadcastable ones.
		
	\end{proof}
	
	The Theorem 2 is most intuitive when inspected in the qubit case. The geometric form of a set of commuting mixed states of qubit states in the Bloch sphere is a line segment passing through the center of the sphere. However, any disk in the Bloch sphere is a maskable set, so any broadcastable set is a proper subset of some maskable sets. As the result, the resources of the maskable set are far more abundant than those of the broadcastable set.
	
	Finally, we comment on the existence of high dimensional maskable set. We have shown in the Theorem 1 that the maximally maskable set lays on a hyperdisk in $d^2-1$-dimensional Euclidean space, that is, the dimension of a maskable set is possible to be as high as $d^2-2$. Generally speaking, the construction of such a set and its corresponding masker is complicated, and remains an open question in the research area of quantum information masking. Here, we takes a step forward by constructing a qudit isometry in even dimension $d=2n$, which masks a $d^2-d$-dimensional set. Consider a group of mappings from $\mathcal{H}^A$ to $\mathcal{H}^A \times \mathcal{H}^B$, acting on the qudit computational bases, $\{\ket{0}, \ket{1}, \ldots, \ket{d-1}\}$, as:
	\begin{align}
		\ket{0} \to \ket{0}^A\ket{0}^B &\to \hspace{29pt} \cos\frac{\alpha_1}{2} \ket{0}^A\ket{0}^B + \sin\frac{\alpha_1}{2} \ket{1}^A\ket{1}^B, \nonumber\\
		\ket{1} \to \ket{1}^A\ket{0}^B &\to e^{-i\theta_1}\left(\sin\frac{\alpha_1}{2} \ket{0}^A\ket{0}^B - \cos\frac{\alpha_1}{2} \ket{1}^A\ket{1}^B\right), \nonumber\\
		\ket{2} \to \ket{2}^A\ket{0}^B &\to \hspace{29pt} \cos\frac{\alpha_2}{2} \ket{2}^A\ket{2}^B + \sin\frac{\alpha_2}{2} \ket{3}^A\ket{3}^B, \nonumber\\
		\ket{3} \to \ket{3}^A\ket{0}^B &\to e^{-i\theta_2}\left(\sin\frac{\alpha_2}{2} \ket{2}^A\ket{2}^B - \cos\frac{\alpha_2}{2} \ket{3}^A\ket{3}^B\right), \nonumber\\
		&\vdots \label{eq:hdmasker}\\
		\ket{d-2} \to \ket{d-2}^A\ket{0}^B &\to \hspace{29pt} \cos\frac{\alpha_n}{2} \ket{d-2}^A\ket{d-2}^B + \sin\frac{\alpha_n}{2} \ket{d-1}^A\ket{d-1}^B, \nonumber\\
		\ket{d-1} \to \ket{d-1}^A\ket{0}^B &\to e^{-i\theta_n}\left(\sin\frac{\alpha_n}{2} \ket{d-2}^A\ket{d-2}^B - \cos\frac{\alpha_n}{2} \ket{d-1}^A\ket{d-1}^B\right). \nonumber
	\end{align}
	The above mappings can be dilated to an isometry. Furthermore, let $\mathcal{D}=\{D_{i}|1\leqslant i\leqslant n\},\, \mathcal{F}=\{F_{jk}|1\leqslant j<k\leqslant n\}$, we can verify that for two groups of fixed real numbers $\{p_1, p_2, \ldots, p_n\}, \{c_1, c_2, \ldots, c_n\}$ with $p_i \geq0, \sum_{i=1}^n p_i=1$, the isometry always masks the following quantum states:
	\begin{align}
		\rho(\mathcal{D}, \mathcal{F}) &= \begin{pmatrix}
			p_1 \mathbf{D}_1 & \mathbf{F}_{12} & \mathbf{F}_{13} & \cdots & \mathbf{F}_{1n} \\
			\mathbf{F}_{12}^\dagger & p_2 \mathbf{D}_2 & \mathbf{F}_{23} & \cdots & \mathbf{F}_{2n} \\
			\mathbf{F}_{13}^\dagger & \mathbf{F}_{23}^\dagger & p_3 \mathbf{D}_3 & \cdots & \mathbf{F}_{3n} \\
			\vdots & \vdots & \vdots & \ddots & \vdots \\
			\mathbf{F}_{1n}^\dagger & \mathbf{F}_{2n}^\dagger & \mathbf{F}_{3n}^\dagger & \cdots & p_n \mathbf{D}_n
		\end{pmatrix},
	\end{align}
	where the $2\times2$ matrices on the diagonal are defined as $\mathbf{D}_i=\frac{1}{2} (I_2+x_i \sigma_x +y_i \sigma_y +z_i \sigma_z)$, with $x_i \sin\alpha_i \cos\theta_i + y_i \sin\alpha_i \sin\theta_i + z_i \cos\alpha_i =c_i$, and the choices for the elements of the $2\times2$ off-diagonal matrices $F_{jk}$ are completely arbitrary, provided that they result in a valid density matrix. To see this more clearly, we directly calculate the marginal states after masking, which are found to be:
	\begin{align}
		\rho_A=\rho_B &=\frac{1}{2}\begin{pmatrix}
			p_1(I_2+c_1\sigma_z) & 0 & 0 & \cdots & 0 \\
			0 & p_2(I_2+c_2\sigma_z) & 0 & \cdots & 0 \\
			0 & 0 & p_3(I_2+c_3\sigma_z) & \cdots & 0 \\
			\vdots & \vdots & \vdots & \ddots & \vdots \\
			0 & 0 & 0 & \cdots & p_n(I_2+c_n\sigma_z)
		\end{pmatrix}.
	\end{align}
	They only depends on $p_i$ and $c_i$, and is irrelevant of the exact forms of $\mathcal{D}$ and $\mathcal{F}$. So, the dimension of the maskable set under this construction is $2|\mathcal{D}|+8|\mathcal{F}| =2n+8(n-1+1)(n-1)/2 =4n^2-2n =d^2-d$. This completes the proof that there exists qudit maskable set that have nonzero measure on $d^2-d$-dimensional Euclidean space. 
	
	\clearpage

	\section{Experimental implementations}
	
	In the experimental part of the Supplementary Material, we first give a detailed account for the relation between the photon fusion gate and the photonic masking machine. Next, using this relation, we show how the extensibility of our masking machine, that how it can be exploited to mask qudit states, and how a qubit generated from practical single-photon source can be masked using a coherent state produced by a pulsed laser. Finally, we discuss a protocol and the experimental result of a masking-based secret sharing, where a secret image is distributed among three recipients. The reconstruction uses no classical communication between the sender and the receivers, but only classical communication between the receivers about the marginal states of their received qubit.
	
	\subsection{The relation between the fusion gate and the masking machine}
	
\begin{figure}[htbp]
		\centering
		\includegraphics[width = .25 \textwidth]{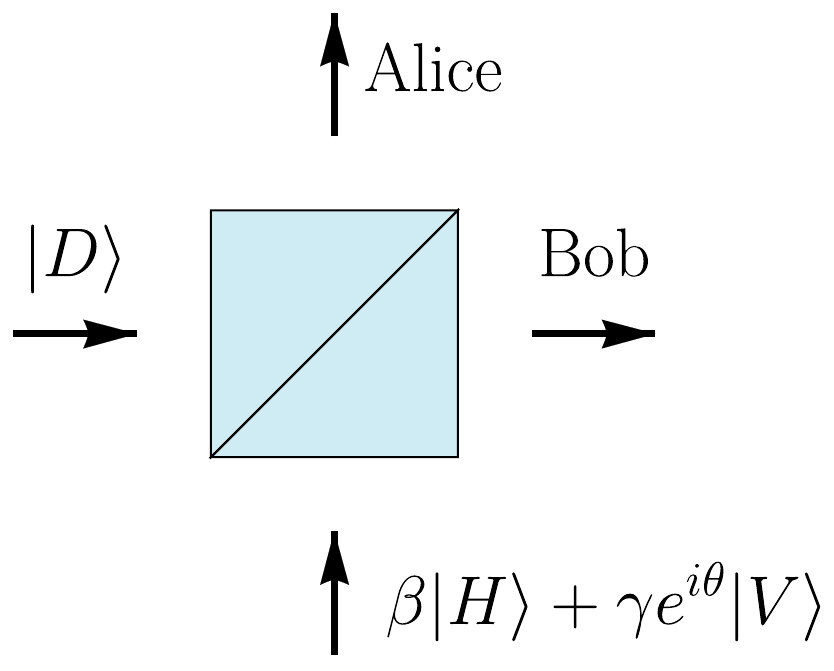}
		\caption{Schematic illustration of the photon fusion gate.}
		\label{figs:fuse}
	\end{figure}

	Here, we give a detailed analysis of the relation between the fusion gate\,\cite{Browne05} and the masking operator $\mathcal{U}_0^0$ in the picture of second quantization. The schematic illustration of the photon fusion gate is shown in Fig.\,\ref{figs:fuse}, where the wavefunction of two photons interfere on a polarizing beam splitter (PBS). 
	The effective surface of the PBS is denoted by the anti-diagonal line crossing the center of the PBS. Regardless of the original direction of propagattion, a photon will be reflected by the PBS if it has vertical polarization (denoted by $\ket{V}$), and will pass through the PBS if its direction of polarization is horizontal (denoted by $\ket{H}$).
	
	A general polarization state $\ket{\psi^A}$ of a single photon can be spanned on the horizontal-vertical basis, namely, $\ket{\psi^A}=\beta \ket{H} + \gamma e^{i\phi} \ket{V}$, with $\beta, \gamma \text{~and~} \phi$ being real numbers and $\beta^2+\gamma^2=1$. For simplicity, in this section we only consider the case of pure states--the situation for mixed states can be proved immediately using the linearity of quantum mechanics.
	In the masking machine (cf. Fig.\,1 in the main text), this photon will interfere with another photon with fixed polarization of $\ket{D}=(\ket{H}+\ket{V})/\sqrt{2}$ on the surface of the PBS. After the PBS gate, the wavefunction propagating from down to is sent to Alice, and the one propagating from left to right is sent to Bob. Moreover, only the photon detection events that result in one photon detected in Alice's site (A) and one in Bob's site (B) are registered as coincidence counting, and the other events are discarded.
	
	To address the quantum interference on the PBS, we introduce the photon creation operator $a_\mu^{\nu\dagger}$ of the photon, which denotes the creation of a photon with polarization of $\mu\in\{H, V\}$ propagating toward the site $\nu\in\{A, B\}$. 
	The conversion of the creation operators of the input photons on the fusion gate reads\,\cite{yhshih}:
	\begin{align}
		\label{eq:modconv}
		a_H^{A\dagger} \to a_H^{A\dagger}, ~~~a_V^{A\dagger} \to ia_V^{B\dagger}, ~~~a_H^{B\dagger} \to a_H^{B\dagger}, ~~~a_V^{B\dagger} \to ia_V^{A\dagger}.
	\end{align}
	
	Using the creation operators, the two-photon input state can be expressed as:
	\begin{align}
		\ket{\psi^A} \otimes \ket{D} &= (\beta \ket{H} + \gamma e^{i\phi} \ket{V})\otimes \ket{D} = (\beta a_H^{A\dagger} + i\gamma e^{i\phi} a_V^{A\dagger}) \,\frac{1}{\sqrt{2}} (a_H^{B\dagger} + a_V^{B\dagger}) \ket{0^A0^B},
	\end{align}
	where $\ket{0^A 0^B}$ denotes the vacuum state. By substituting (\ref{eq:modconv}) into the above equation, we acquire the form of the two-photon state after the fusion gate:
	\begin{align}
		\ket{\psi^A} \otimes \ket{D} \overset{\text{Fusion}}{\longrightarrow} \ket{\psi^{AB}} &= (\beta a_H^{A\dagger} + i\gamma e^{i\phi} a_V^{B\dagger}) \frac{1}{\sqrt{2}} (a_H^{B\dagger} + i a_V^{A\dagger}) \ket{0^A 0^B} \nonumber \\
		&= \left(\frac{\beta a_H^{A\dagger} a_H^{B\dagger} -\gamma e^{i\phi} a_V^{A\dagger} a_V^{B\dagger}}{\sqrt{2}} + i\frac{\beta a_H^{A\dagger} a_V^{A\dagger} + \gamma e^{i\phi} a_H^{B\dagger} a_V^{B\dagger}}{\sqrt{2}}\right) \ket{0^A 0^B}.
	\end{align}
	The second fraction in the parenthesis, however, does not result in coincidence counting. Consequently, the effective wavefunction after the PBS that reaches coincidence counting, after normalization, reads:
	\begin{align}
		\ket{\psi^{AB}}=\beta \ket{H^A H^B} - \gamma e^{i\phi} \ket{V^A V^B}.
	\end{align}
	Observe that the fusion gate effectively implements an entangling projection $\ket{HH}\bra{HH}-\ket{VV}\bra{VV}$ on the input two-photon state. This conclusion is first drawn in \cite{lmduan06}, and here the form of the projector is slightly different due to the extra $\pi/2$ phase the photons picked up upon reflection\,\cite{yhshih}.
	
	By checking the two marginal density operators of the bipartite system, we find,
	\begin{align}
		\mathrm{Tr}_A\left(\ket{\psi^{AB}}\bra{\psi^{AB}}\right) &= \mathrm{Tr}_B\left(\ket{\psi^{AB}}\bra{\psi^{AB}}\right) = \begin{pmatrix}
			\beta^2 & 0 \\
			0 & \gamma^2
		\end{pmatrix}.
		\label{eq:dmreduced}
	\end{align}
	We see that the parameter $\phi$ has ceased to appear in (\ref{eq:dmreduced}), which means that the class of quantum states with the same $\beta$, $\gamma$ and different phases $\phi$ are masked. These states lays on a latitudinal parallel on the Bloch sphere. Following the definition of the masker, we conclude that the fusion gate realizes $\mathcal{U}_0^0$.
	
	Furthermore, we explicitly give the form of the masking isometry $\mathcal{U}$ from the form of mode conversion taking place in the fusion gate. Because the fusion gate and the Hadamard gate acting on the auxiliary photon link the input and the output states as $\begin{pmatrix}\beta \\ \gamma\end{pmatrix} \otimes \begin{pmatrix}1 \\ 0\end{pmatrix} \overset{\text{Fusion}}{\longrightarrow} \begin{pmatrix}\beta \\ 0 \\ 0 \\ -\gamma\end{pmatrix}$, up to a renormalization of the final state, we can choose
	\begin{align}
		\mathcal{U}_0^0=\begin{pmatrix}1 & 0 & 0 & 0 \\ 0 & 1 & 0 & 0 \\ 0 & 0 & 0 & -1 \\ 0 & 0 & -1 & 0\end{pmatrix}.
	\end{align}
	Note that the from of the second and the fourth columns of the $\mathcal{U}_0^0$ can be freely chosen, as long as the value assignments preserve unitarity of the evolution. Finally, because
	that a $SU(2)$ rotation $\mathcal{R}_\alpha^\theta = \begin{pmatrix}\cos (\alpha/2) & e^{-i\theta}\sin (\alpha/2) \\ \sin (\alpha/2) & -e^{-i\theta}\cos (\alpha/2) \end{pmatrix}$ rotates the edge of the disk $\mathcal{D}_\alpha^\theta (\rho_0)$ to a latitudinal circle on Bloch sphere, we can construct the masking isometry $\mathcal{U}_\alpha^\theta$ as is given in (\ref{eqs:uat}) by applying the $SU(2)$ rotation before the masking gate. By doing this, the final form of the mapping becomes:
	\begin{align}
		\mathcal{U} &= \mathrm{CNOT}\,(\mathcal{R}(\alpha, \theta)\otimes I_2) = \begin{pmatrix}\cos(\alpha/2) & 0 & e^{-i\theta}\sin(\alpha/2) & 0 \\ 0 & \cos(\alpha/2) & 0 & e^{-i\theta}\sin(\alpha/2) \\ 0 & \sin(\alpha/2) & 0 & -e^{-i\theta}\cos(\alpha/2) \\ \sin(\alpha/2) & 0 & -e^{-i\theta}\cos(\alpha/2) & 0\end{pmatrix},
	\end{align}
	which is in accord with the masking isometry defined in (\ref{eqs:uat}). 
	
	\subsection{Extensibility of the masking protocol for qudit}
	
\begin{figure}[htbp]
		\centering
		\includegraphics[width=.4 \textwidth]{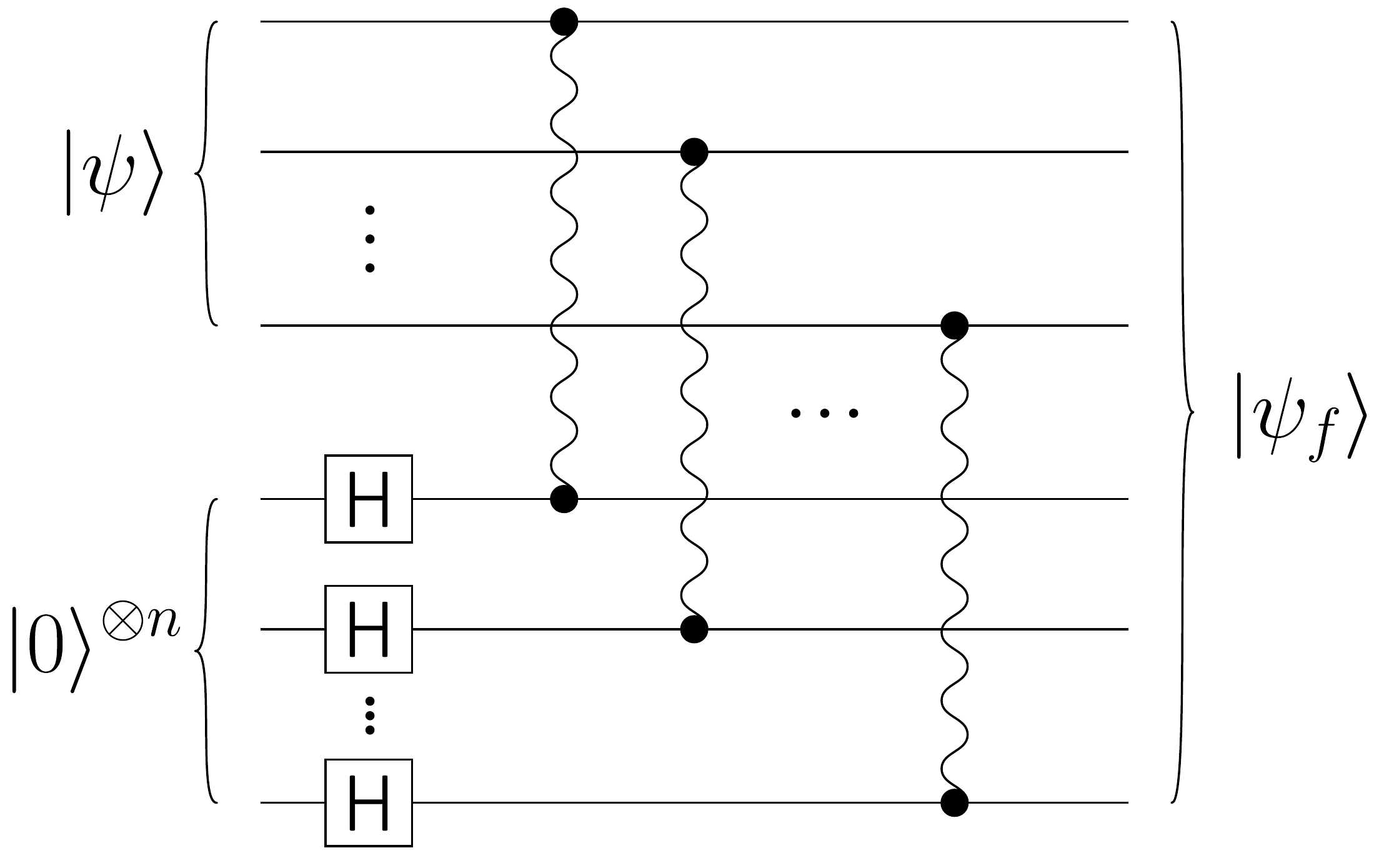}
		\caption{Applying the masking machine on qudit states. The tilde line denotes the fusion gate.}
		\label{fig:qudit}
	\end{figure}

	The polarization degree of freedom offers a natural photonic qubit state suitable for quantum information processing. In order to realize the masking isometry in higher dimensions, we can introduce more photon pairs to synthesize a composite qudit from some qubits. The quantum circuit for masking higher-dimensional states is shown in Fig.\,\ref{fig:qudit}. Below, we explain the scheme in detail. 
	
	A composite qudit $\ket{\psi}=c_0\ket{0}+c_1e^{i\phi_1}\ket{1} +\cdots+c_{d-1}e^{i\phi_{d-1}}\ket{d-1}$ with the dimension of $d\leq2^n$ can be encoded digit-wise on the horizontal ($H$) and vertical ($V$) polarizations of $n$ photons $\ket{\cdot}^1 \ket{\cdot}^2 \cdots \ket{\cdot}^n$, and the correspondence between the qudit and the joint states of the photon polarizations reads 
	\begin{align*}
		\ket{0}&\leftrightarrow \ket{HH\ldots HH}, \\
		\ket{1}&\leftrightarrow \ket{HH\ldots HV}, \\
		\ket{2}&\leftrightarrow \ket{HH\ldots VH}, \\ 
		\ket{3}&\leftrightarrow \ket{HH\ldots VV}, \\ 
		&\cdots \\
		\ket{2^n-1}&\leftrightarrow \ket{VV\ldots VV}.
	\end{align*}
	Henceforth, for simplicity, we have omitted the superscript of photon indexing. To mask the target qudit we need an auxiliary qudit, again composed of $n$ photons, and set at the polarization state $\ket{D}^{\otimes n}$ with $\ket{D}=(\ket{H}+\ket{V})/\sqrt{2}$. Each of the carrier photon then undergoes a fusion operation with an auxiliary photon on a PBS.
	We write down the initial state of the two-qudit system before the fusion gate on the polarization bases, namely,
	\begin{align}
		\ket{\psi}\otimes\ket{D}^{\otimes n} = \left(\sum_{k=0}^{d-1} c_k e^{i\phi_k} \ket{k}\right)\ket{D}^{\otimes n}.
	\end{align}
	
	The photon fusion gate effectively conducts a postselection so that one and only one photon emerges from each of the output modes. Consequently, for each fusion gate, the two output photons necessarily have the same polarization. In this case, because the inputs of each fusion gate are still single photon, we simply project the corresponding two-photon states on the $\ket{HH}\bra{HH}-\ket{VV}\bra{VV}$ basis (the minus sign before the $\ket{VV}\bra{VV}$ terms is due to the extra $\pi/2$ phase the photons picked up upon reflection) to acquire the evolutions of the calculation bases, which read:
	\begin{align}
		\ket{0}\otimes\ket{D}^{\otimes n} &=\ket{HH\ldots HH}\left\vert DD\ldots DD\right\rangle \nonumber\\
		&\overset{\text{Fusion}}{\longrightarrow} \phantom{(-1)\times} 2^{-n/2}\ket{HH\ldots HH}\ket{HH\ldots HH} = 2^{-n/2}\ket{0}\ket{0}, \nonumber\\
		\ket{1}\otimes\ket{D}^{\otimes n} &=\ket{HH\ldots HV}\left\vert DD\ldots DD\right\rangle \nonumber\\
		&\overset{\text{Fusion}}{\longrightarrow} (-1)\times 2^{-n/2}\ket{HH\ldots HV}\ket{HH\ldots HV} = 2^{-n/2}\ket{1}\ket{1}, \nonumber\\
		\ket{2}\otimes\ket{D}^{\otimes n} &=\ket{HH\ldots VH}\left\vert DD\ldots DD\right\rangle \nonumber\\
		&\overset{\text{Fusion}}{\longrightarrow} (-1)\times 2^{-n/2}\ket{HH\ldots VH}\ket{HH\ldots VH} = 2^{-n/2}\ket{2}\ket{2}, \nonumber\\
		\ket{3}\otimes\ket{D}^{\otimes n} &=\ket{HH\ldots VV}\left\vert DD\ldots DD\right\rangle \nonumber\\
		&\overset{\text{Fusion}}{\longrightarrow} \phantom{(-1)\times} 2^{-n/2}\ket{HH\ldots VV}\ket{HH\ldots VV} = 2^{-n/2}\ket{3}\ket{3}, \nonumber\\
		&\vdots \nonumber\\
		\ket{k}\otimes\ket{D}^{\otimes n} &=\ket{k}\left\vert DD\ldots DD\right\rangle \overset{\text{Fusion}}{\longrightarrow} (-1)^{\mathcal{P}(k)}\times 2^{-n/2}\ket{k}\ket{k}, \\
		&\vdots \nonumber\\
		\ket{d}\otimes\ket{D}^{\otimes n} &=\ket{d}\left\vert DD\ldots DD\right\rangle \overset{\text{Fusion}}{\longrightarrow} (-1)^{\mathcal{P}(d)}\times 2^{-n/2}\ket{d}\ket{d}. \nonumber
	\end{align}
	Here, we have defined $(-1)^{\mathcal{P}(k)}$ as the parity of $k$. In other words, $\mathcal{P}(k)$ is the count of 1's in the binary form of $k$. The result is based on the observation, that every vertically-polarized photon encoding the computational basis $\ket{k}$ of the composite qudit contributes a $\pi$-phase when affected by a fusion gate. Consequently, the quantum state $\ket{\psi_f}$ resulted by fusing pairwise the carrier and auxiliary photons reads:
	\begin{align}
		\ket{\psi}\otimes\ket{D}^{\otimes n} \overset{\text{Fusion}}{\longrightarrow} \ket{\psi_f} = \sum_{k=0}^{d-1} (-1)^{\mathcal{P}(k)} c_k e^{i\phi_k} \ket{k}\ket{k}.
	\end{align}
	Further, the marginal state for Alice, who receives only one of the output composite qudit, reads:
	\begin{align*}
		\rho_f^\text{A} &= \mathrm{Tr}^{B}[\ket{\psi_f}\bra{\psi_f}] = \sum_{k=0}^{d-1} |c_k|^2 \ket{k}\bra{k},
	\end{align*}
	and, by symmetry, the marginal state for Bob is $\rho_f^\text{B} = \rho_f^\text{A}$.
	
	We see that the relative phases $e^{i\phi_k}$ in the initial state $\ket{\psi}$ have no observable effect on the two receiving party's marginal state. By definition of quantum information masking, we have masked a proper subset of a qudit Hilbert space. In conclusion, by introducing additional photons, quantum informatiom masking can be extended to arbitrary high dimension using our setup, and the masking isometry acts as $\ket{\psi}=\sum_k c_k \ket{k} \to \ket{\psi_f}=\sum_k (-1)^{\mathcal{P}(k)}c_k \ket{kk}$, with $\mathcal{P}(k)$ being the total number of 1's in the   binary representation of $k$.
	
	\subsection{Extensibility of the masking protocol for different photon sources}
	
	In the main article, we demonstrated the photonic quantum information masking using the parametric photons from pumping the nonlinear crystals. However, the scope of the masking machine is not limited by the specific photon source. Since the fusion gate only requires overlapping and indistinguishability of the two-photon wavefunction on the PBS to implement the mode conversion, the masking machine may also accept other types of photon sources as the input. Here, we show that the weak coherent light (e.g., attenuated laser beam) can be exploited to mask the quantum information carried by single photons.  
	
	\begin{figure}[htbp]
		\centering
		\includegraphics[width=.28 \textwidth]{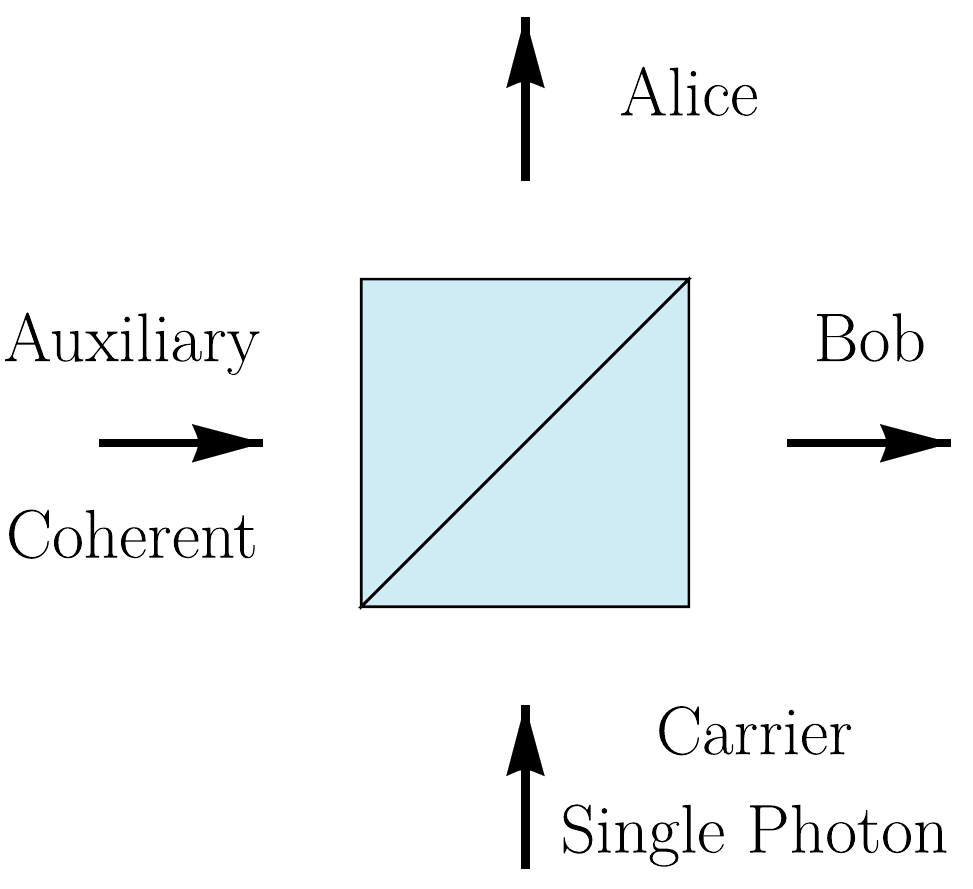}
		\caption{Schematic illustration of the fusion gate with a single-photon source and a coherent state as the inputs.}
		\label{fig:fusec}
	\end{figure}
	
	The schematic illustration of the scenario is shown in Fig.\,\ref{fig:fusec}. Specifically, we are interested to mask a carrier photonic qubit, which is generated by a periodically pumped  semi-deterministic single-photon source, using an auxiliary, pulsed laser which have the same spectral profile as the carrier photon, and the same repetition rate, so a single-photon always meets a coherent pulse at the PBS. 
	In this case, because the input modes of the fusion gate are no longer single-photons, it is not possible to directly project the input states on $\ket{HH}\bra{HH}-\ket{VV}\bra{VV}$ to extract the output state, and the situation must be treated in the second quantization picture. We again span the polarization state of the carrier photon on the horizontal-vertical basis, that is, $\ket{\psi^A}=\beta\ket{H}+\gamma e^{i\phi}\ket{V}$, and denote as $p$ the efficiency of the single-photon source to successfully generate a photon. Then, the photon state on the carrier side can be expressed as:
	\begin{align}
		\ket{\psi^A}=\left[\sqrt{1-p}+\sqrt{p}(\beta a_H^{A\dagger}+\gamma e^{i\phi} a_V^{A\dagger})\right]\ket{0^A}.
	\end{align}
	On the other hand, the polarization of photons on the auxiliary side is still set to $\ket{D}$. The second quantization form of the input coherent state reads:
	\begin{align}
		\ket{\alpha}=\sum_{k=0}^{+\infty}\frac{1}{k!}\left(\alpha\, a_D^{B\dagger}\right)^k\ket{0^B}=\sum_{k=0}^{+\infty}\frac{\alpha^k}{k!}\left(\frac{a_H^{B\dagger}+a_V^{B\dagger}}{\sqrt{2}}\right)^k\ket{0^B}.
	\end{align}
	Also, that the coherent laser is weak implies $|\alpha|\ll1$, so we make use of the Maclaurin series expansion up to the second order of $\alpha$ to find: 
	\begin{align}
		\ket{\alpha} \approx \left[1+\frac{\alpha}{\sqrt{2}}\left(a_H^{B\dagger}+a_V^{B\dagger}\right)+\frac{\alpha^2}{4}\left(a_H^{B\dagger}a_H^{B\dagger}+2a_H^{B\dagger}a_V^{B\dagger}+a_V^{B\dagger}a_V^{B\dagger}\right)\right]\ket{0^B}.
	\end{align}
	
	We are now in the position to analyze the efficacy of the masking machine for the scenario of single-photon plus coherent laser input. Applying the mode conversion rule (\ref{eq:modconv}) on the joint input state yields
	\begin{align}
		\ket{\psi^A}\otimes\ket{\alpha} =& \left[\sqrt{1-p}+\sqrt{p}(\beta a_H^{A\dagger}+\gamma e^{i\phi} a_V^{A\dagger})\right] \left[1+\frac{\alpha}{\sqrt{2}}\left(a_H^{B\dagger}+a_V^{B\dagger}\right)+\frac{\alpha^2}{4}\left(a_H^{B\dagger}a_H^{B\dagger}+2a_H^{B\dagger}a_V^{B\dagger}+a_V^{B\dagger}a_V^{B\dagger}\right)\right] \ket{0^A0^B}\nonumber\\ \overset{\text{Fusion}}{\longrightarrow}& \left[\sqrt{1-p}+\sqrt{p}(\beta a_H^{A\dagger}+i\gamma e^{i\phi} a_V^{B\dagger})\right] \left[1+\frac{\alpha}{\sqrt{2}}\left(a_H^{B\dagger}+ia_V^{A\dagger}\right)+\frac{\alpha^2}{4}\left(a_H^{B\dagger}a_H^{B\dagger}+2ia_H^{B\dagger}a_V^{A\dagger}-a_V^{A\dagger}a_V^{A\dagger}\right)\right] \ket{0^A0^B}.
	\end{align}
	However, the system subjects to the postselection from coincidence counting, so the final recorded state only contains the terms which have creation operators for both Alice and Bob. By ignoring the vacuum state, expanding the product and discarding the irrelevant terms that does not give coincidence counting, we find:
	\begin{align}
		\ket{\psi^A}\otimes\ket{\alpha} \overset{\text{Fusion}}{\longrightarrow}& \left[\sqrt{\frac{p}{2}}\,\alpha(\beta a_H^{A\dagger}+i\gamma e^{i\phi}a_V^{B\dagger}) \left(a_H^{B\dagger}+ia_V^{A\dagger}\right) + \frac{\alpha^2\sqrt{1-p}}{4} \left(a_H^{B\dagger}a_H^{B\dagger}+2ia_H^{B\dagger}a_V^{A\dagger}-a_V^{A\dagger}a_V^{A\dagger}\right)\right]\ket{0^A0^B} \nonumber\\
		\overset{\text{coin.}}{\longrightarrow}& \left[\sqrt{\frac{p}{2}}\,\alpha(\beta a_H^{A\dagger} a_H^{B\dagger} -\gamma e^{i\phi}a_V^{A\dagger}a_V^{B\dagger}) + \frac{i\alpha^2\sqrt{1-p}}{2}a_H^{B\dagger}a_V^{A\dagger} \right]\ket{0^A0^B}.
	\end{align}
	So, the final state $\ket{\psi^{AB}}$ detected by coincidence counting can be obtained by applying the creation operators on the vacuum state, and switch back to the notation of first quantization. It reads:
	\begin{align}
		\ket{\psi^{AB}} &= (\beta\ket{HH}-\gamma e^{i\phi}\ket{VV}) + i\sqrt{\frac{1-p}{2p}}a\ket{HV},
	\end{align}
	in which the first term is the desired outcome of the masking machine, and the second term is resulted by the multi-photon terms in the coherent state. The normalization is made for the first term. We see that the noise term vanish at the limit $p\gg a$, i.e., when the mean photon numbers of the coherent state per pulse is much less than the efficiency of the single-photon source. The condition is well attainable with current state-of-the-art deterministic single-photon sources, where some realizations with $p>0.7$ have been reported. Our calculation shows that the photonic quantum information masking machine is also exploitable for the broader class of photon sources, and may find its role in future single-photon source-based quantum information and quantum communication scenarios.

	\subsection{Image reconstruction based on qubit masking}
	
	In the main text, we have proposed a protocol for a tripartite sharing of a secret qubit state. The sender, Alice, distributes a secret quantum state to three recipients, Bobs, who can only use their marginal state to restrict the masked state onto a disk, and have to cooperate together to decode the secret. Precisely, Alice should mask the secret qubit state $\rho_0$ with three different maskers $\mathcal{U}_{(i)}$ and sends one of the qubits from the resulted $\rho_i^{AB}$ to $\text{Bob}_i$, respectively. The Bobs are informed \textit{a priori} the form of the maskers $\mathcal{U}_{(i)}$. However, from the received marginal state, $\text{Bob}_i$ can only restrict the masked state onto a disk. It is when all Bobs cooperate together and compare their result of the marginal state (using only classical communication) that they can unlock the secret. Upon comparison, the three disks will intersect at a single point in the Bloch sphere, which reveals the concealed information $\rho_0$. 
	Our protocol also offers some security against eavesdropping providing that the masker and the marginal state are not simultaneously divulged: fabrication of either of the two components in masking may eventuate in reconstruction of a non-physical state that locates outside the Bloch sphere, which exposes the eavesdropper.
	
	Furthermore, utilizing the homomorphism between the Bloch representation of a qubit and the hue-saturation-luminosity color space, we can use a qubit state to encode a colored pixel, and share a secret picture pixel-wise between three recipients.
	In our protocol of image sharing, the color of each pixel is determined by three float numbers interpreted by the receivers, ``Bobs''. For this course, a correspondence between the string of numbers decoded from Bobs' quantum states and the reconstructed color has to be established. Here, this is achieved by resorting to the hue-saturation-luminosity (HSL) representation of a color. 
	
	We recall the geometric representation for any mixed qubit state in the Bloch sphere, that is,
	$$\rho(\vec{r}) = \rho(x, y, z) = \frac{1}{2} \left(I_2+x\sigma_x+y\sigma_y+z\sigma_z\right). $$
	
	For every point in the Bloch sphere, we define the three color parameters, viz., hue, saturation and luminosity, as
	\begin{align*}
		h &= \frac{1}{2} + \frac{\arctan(x, y)}{2\pi}, \\
		s &= \frac{x^2+y^2}{1-z^2}, \\
		l &= \frac{1+z}{2}.
	\end{align*}
	respectively, where $\arctan(x, y)\in[-\pi, \pi)$ is the arc tangent of $y/x$, taking into account which quadrant the point $(x, y)$ is in. The value of hue is undefined at the points $(x, y, z)$ with $x=y=0$, and the saturation is undefined at $z=\pm1$. These singularities will not cause confusion to the final rendition of colors, which is described using the red, green and blue (RGB) values of the colors. The RGB and HSL values of a color is linked by
	\begin{align*}
		\{r, g&, b\} = \{f(0), f(8), f(4)\}, \\
		\text{with}~~~ f(n) &= l-s\min(l, 1-l) \max[-1, \min(k-3, 9-k, 1)], \\
		\text{and}~~~~ k&=\left(n+\frac{6h}{\pi}\right) \mod 12.
	\end{align*}
	\begin{figure}[htbp]
		\centering
		\includegraphics[width = .24 \textwidth]{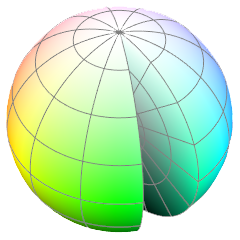}
		\caption{The one-on-one correspondence between possible colors and quantum state's location in the Bloch sphere.}
		\label{figs:hsl}
	\end{figure}
	
	The bijection of the RGB colors and the quantum state's location in the Bloch sphere is illustrated in Fig\,\ref{figs:hsl}. The singularities on the $z$-axis resolve because they represent grayscale colors can be solely described by the luminosity. 
	In our experiment, for every pixel to be transmitted, Alice encodes the message $(x_0, y_0, z_0)$ into the quantum state $\rho_0$. By applying the set of maskers $\mathcal{U}_{\pi/2}^0, \mathcal{U}_{\pi/2}^{\pi/2}$ and $\mathcal{U}_0^0$ on $\rho_0$, and sending one of the resulted qubits to each Bob, she effectively prepares the marginal states for the three Bobs to be $\begin{pmatrix}(1+x_0)/2 & 0 \\ 0 & (1-x_0)/2\end{pmatrix}$, $\begin{pmatrix}(1+y_0)/2 & 0 \\ 0 & (1-y_0)/2\end{pmatrix}$ and $\begin{pmatrix}(1+z_0)/2 & 0 \\ 0 & (1-z_0)/2\end{pmatrix}$, respectively. 
	From the form of the masker and the marginal state, each of the three Bobs can determine a disk in the Bloch sphere that the original state must belong to, and the three disks lays on three mutually orthogonal planes so they necessarily intercept at one point in the Bloch sphere. The intercepting point in turn represents the color information of the transmitted pixel.
	
	\begin{figure}
		\centering
		\includegraphics[width=.6 \textwidth]{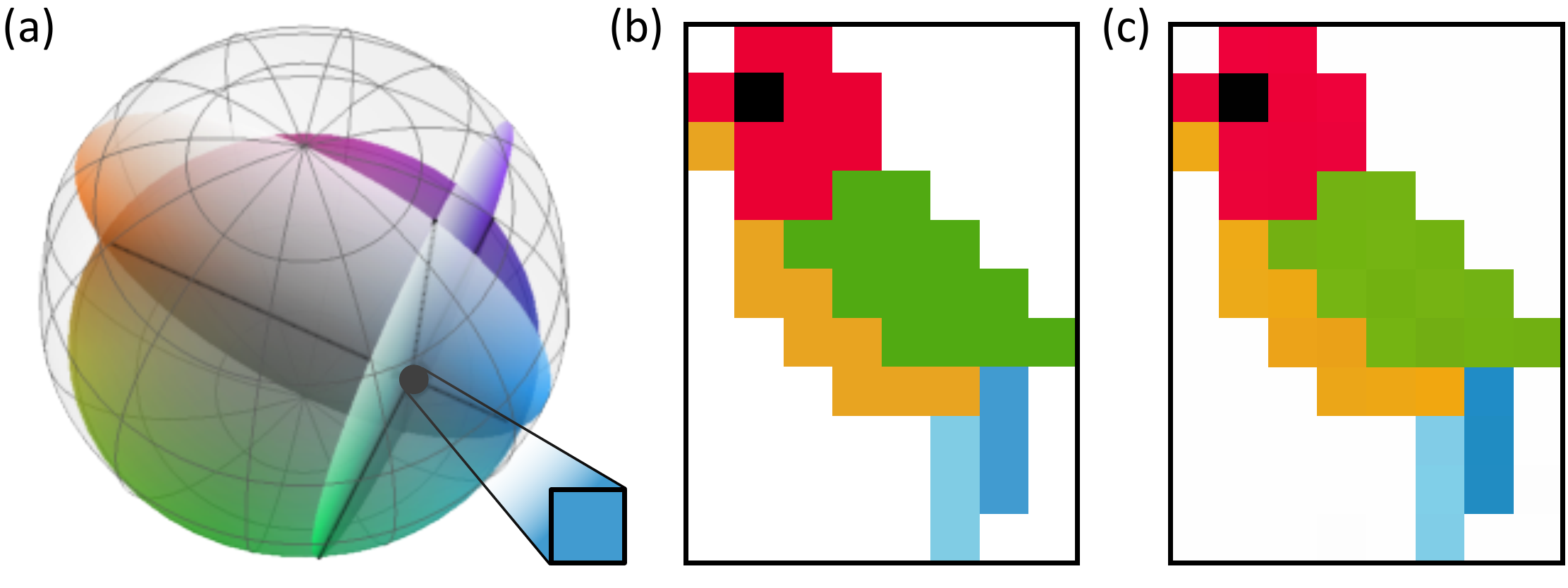}
		\caption{Tripartite secret sharing of an image by masking of mixed states. (a) From their corresponding reduced states and the masker information informed by Alice, every Bob independently interpret the possible disk that contains the masked states. The original state is pinned at the intersection of the three disks, and the location uniquely determines a color using the hue-saturation-luminosity representation. (b) A pixel art of a parrot to be shared. (c) The reconstructed image from the input of three Bobs.}
		\label{fig:share}
	\end{figure}
	
	In our experiment, Alice adopts $\mathcal{U}_{(1)}=\mathcal{U}_{\pi/2}^0,\mathcal{U}_{(2)}=\mathcal{U}_{\pi/2}^{\pi/2}\text{~and~}\mathcal{U}_{(3)}=\mathcal{U}_0^0$ to mask the quantum state $\rho_0$ corresponding to the color of each pixel. The extraction protocol is displayed in Fig.\,\ref{fig:share}a. From the bipartite state resulted by applying $\mathcal{U}_{(i)}$, one particle is sent to $\text{Bob}_i$ and one is kept by Alice. Because the reduced density matrix for Bob after applying the maskers on $\rho_0$ are $(I_2+x\sigma_z)/2,(I_2+y\sigma_z)/2,~\text{and}~(I_2+z\sigma_z)/2$, respectively, this operation effectively grants each Bobs access to one of the float numbers, $x_0,y_0,\text{~and~}z_0$, which constitutes the original state. By repeatedly applying the above procedure, the entire image is split and transmitted. The Bobs can then assemble their deduced parameters to recover $\rho_0$, which corresponds to a certain color of a pixel. The reconstructed image in Fig.\,\ref{fig:share}c reasonably resembles the original one (cf. Fig.\,\ref{fig:share}b), and the correlation between the two colored images, averaging over red-green-blue channels, is calculated to be $99.35\%$.
	
	\let\oldaddcontentsline\addcontentsline
	\renewcommand{\addcontentsline}[3]{}

	\let\addcontentsline\oldaddcontentsline 
	
\end{document}

\end{document}